\documentclass[10pt, conference, letterpaper]{ IEEEtran}
\IEEEoverridecommandlockouts
\usepackage{cite}
\usepackage{amsmath,amssymb,amsfonts}
\usepackage{graphicx}
\usepackage{textcomp}
\usepackage{xcolor}

\usepackage{flushend}

\usepackage{graphicx}
\usepackage{booktabs}
\usepackage{array}
\usepackage{algorithm}
\usepackage{algpseudocode}
\def\BibTeX{{\rm B\kern-.05em{\sc i\kern-.025em b}\kern-.08em
    T\kern-.1667em\lower.7ex\hbox{E}\kern-.125emX}}
\begin{document}

\title{WiDistill: Distilling Large-scale Wi-Fi Datasets with  Trajectory Matching
}

\author{\IEEEauthorblockN{Tiantian Wang}
\IEEEauthorblockA{\textit{School of Software Engineering} \\
\textit{Xi'an Jiaotong University}\\
Xi'an, China \\
tiantianwang@stu.xjtu.edu.cn}
\and
\IEEEauthorblockN{Fei Wang}
\IEEEauthorblockA{\textit{School of Software Engineering} \\
\textit{Xi'an Jiaotong University}\\
Xi'an, China \\
feynmanw@xjtu.edu.cn}
}

\maketitle

\begin{abstract}
Wi-Fi based human activity recognition is a technology with immense potential in home automation, advanced caregiving, and enhanced security systems. It can distinguish human activity in environments with poor lighting and obstructions. However, most current Wi-Fi based human activity recognition methods are data-driven, leading to a continuous increase in the size of datasets. This results in a significant increase in the resources and time required to store and utilize these datasets. To address this issue, we propose WiDistill, a large-scale Wi-Fi datasets distillation method. WiDistill improves the distilled dataset by aligning the parameter trajectories of the distilled data with the recorded expert trajectories. WiDistill significantly reduces the need for the original large-scale Wi-Fi datasets and allows for faster training of models that approximate the performance of the original network, while also demonstrating robust performance in cross-network environments. Extensive experiments on the Widar3.0, XRF55, and MM-Fi datasets demonstrate that WiDistill outperforms other methods. The code can be found in https://github.com/the-sky001/WiDistill.
\end{abstract}

\begin{IEEEkeywords}
wifi sensing, dataset distillation, deep learning
\end{IEEEkeywords}

\section{Introduction}\label{sec:introduction}
\subsection{Background and Motivation}
Wi-Fi based human activity recognition is a critical technology with significant implications for various domains, including home automation\cite{elshafee2012design,doshi2012improving}, advanced caregiving\cite{song2014development}, and enhanced security systems\cite{zhang2020wifi}. By utilizing Wi-Fi CSI signals, this technology can accurately identify human activity even in environments with poor lighting and obstructions, where traditional vision-based systems may fail\cite{wang2019person,wang2019joint,wang2018csi,wang2019can}. The widespread availability of Wi-Fi devices further facilitates the integration and deployment of this technology, making it both practical and accessible\cite{boudlal2023design}. However, current Wi-Fi based human activity recognition methods rely heavily on large-scale data, which leads to excessive resource consumption for data computation  and high cost of transmission and processing. The traditional coreset selection can be considered as a solution method\cite{guo2022deepcore}. By selecting representative samples in the dataset through optimization-based or sampling-based methods, it uses the ensemble of representative samples as a kind of condensation of the original dataset. However,  conventional methods have a low data compression rate. When compressed to a certain degree, the performance of the compressed dataset will be greatly reduced. Additionally, it is poorly adaptable in cross-domain scenarios because the selected compressed dataset may perform well under certain conditions but poorly when applied across different networks or environments.
\begin{figure}
    \centering
    \includegraphics[width=0.9\linewidth]{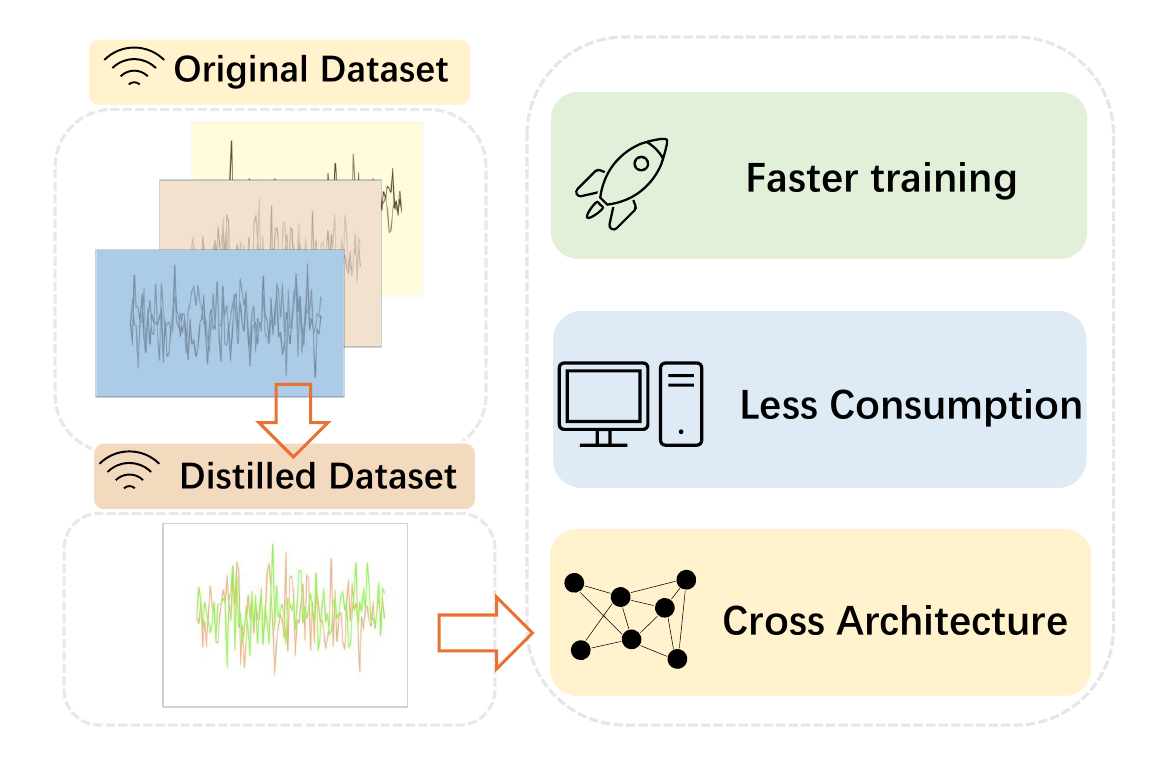}
    \caption{The distilled WiFi dataset will have faster training speed and consume less training time and resources due to the reduced data size, and will also perform better in cross-domain environments due to the more representative nature of the distilled dataset.}
    \label{fig:enter-label}
\end{figure}

Dataset distillation\cite{wang2018dataset} offers a promising solution to these challenges by condensing a given dataset into a synthetic version that retains the essential information of the original data. Training on these synthetic datasets is expected to approximate the performance of training on the original datasets.
\begin{figure*}[h!]
    \centering
    \includegraphics[width=0.9\linewidth]{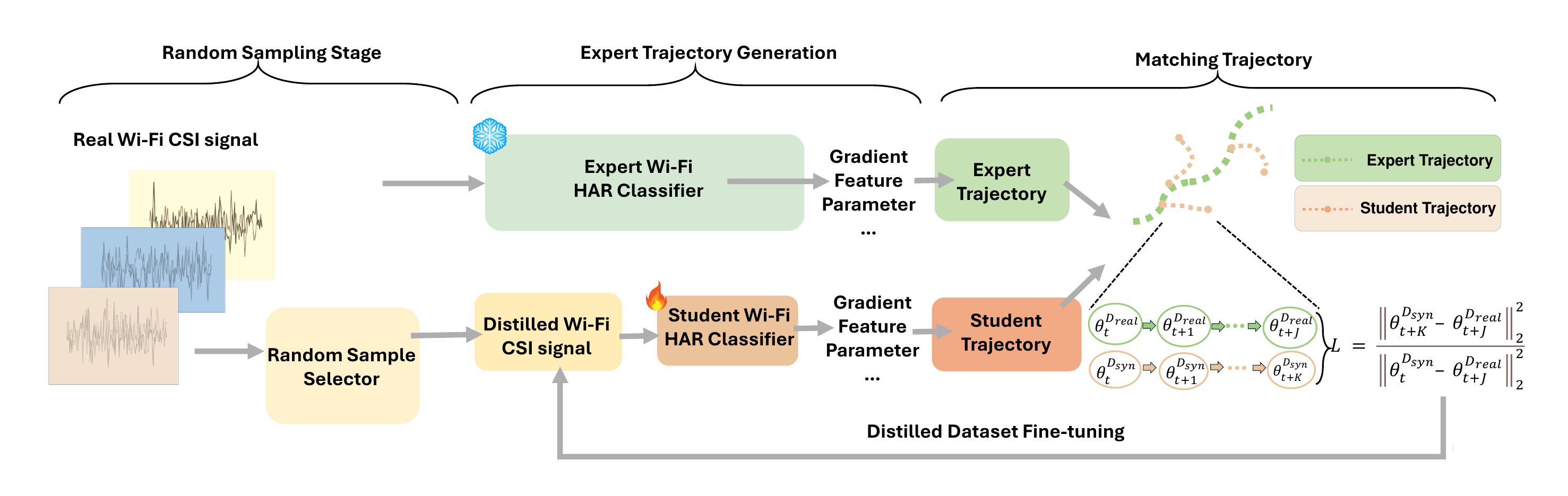}
    \caption{We design WiDistill in three  stages, which are expert trajectory generation, matching trajectory, and
distilled dataset fine-tuning. First, we serve as a prototype for distillation by randomly sampling the original real dataset. Then we train the real dataset completely and record the parameters, features, gradients  as expert trajectory. Then we train the distillation dataset and record the student trajectories, after which we compare the difference between the student trajectories and the expert trajectories at random times and use this to guide the fine-tuning of the distillation dataset.
}
    \label{fig:enter-label}
\end{figure*}
 
 In this paper, we propose WiDistill to distill large-scale Wi-Fi datasets into smaller distilled datasets while maintaining similar performance. To our best knowledge, we are the first to apply dataset distillation to Wi-Fi datasets. Although existing dataset distillation methods have been evaluated on natural image datasets, their effectiveness for distilling Wi-Fi datasets has not been comprehensively understood or verified. Compared to natural images, Wi-Fi CSI signals and their associated analysis tasks exhibit fundamentally different characteristics and technical challenges. Therefore, it is essential to establish a comprehensive benchmark for large-scale Wi-Fi dataset distillation and identify the challenges along with potential solutions. The main contributions are summarized as follows:

 1) In the field of Wi-Fi based human activity recognition, it is common to enhance model performance by increasing the size of datasets, such as expanding the coverage of the dataset through data generation\cite{Korany_Karanam_Cai_Mostofi_2019,hou2024rfboost}. We propose a distillation method for large-scale Wi-Fi datasets (WiDistill), which compresses large-scale Wi-Fi datasets into smaller ones while maintaining similar performance.
 
 2) Although dataset distillation techniques have made significant progress in the field of computer vision, they have yet to be applied in the domain of Wi-Fi human activity recognition. We pioneer the application of a dataset distillation method based on trajectory matching, specifically designed for CSI signals, to distill large-scale WiFi datasets.
 
 3) Experiments on three public datasets demonstrate the effectiveness of cross-modal features, and our proposed WiDistill  outperforms other approaches in terms of recognition accuracy while significantly reducing the dataset size.

\section{Method}\label{sec:method}
\subsection{Overview}

The goal of our research is to generate a synthetic dataset \(D_{\text{syn}}\) whose performance is comparable to that of the real large-scale Wi-Fi dataset \(D_{\text{real}}\) with the same settings. Distillinig the large-scale Wi-Fi dataset is beneficial for lowering computational costs, decreasing training time, and facilitating the edge deployment of Wi-Fi based human activity recognition. To this end, WiDistill employs trajectory matching  to minimize the gap between the trajectories of the distilled dataset and the real dataset, maintaining similar performance while reducing the dataset size. WiDistill consists of three stages: expert trajectory generation, matching trajectory, and distilled dataset fine-tuning.

The expert trajectory generation stage is the core of WiDistill. It involves training a neural network on the complete real dataset and recording various parameters, which we refer to as expert trajectory. These are crucial for the subsequent data distillation process. In the matching trajectory stage, 
We first records the student trajectories generated after running the distilled dataset for a certain period, and then matches them with the expert trajectories. In the distilled dataset fine-tuning stage, we use the distance between the parameters as the loss, applying backpropagation to adjust the distilled dataset, bringing its performance closer to that of the real dataset. Fig. \ref{fig:enter-label} shows the system overview of WiDistill.

\subsection{Wi-Fi based human activity recognition}
Firstly, we introduce the theoretical basis of Wi-Fi based human activity recognition. Channel state information (CSI) reflects the changes in the amplitude and phase of the wireless signal during the signal propagation. In practice, the received signal is a superposition of signals from multiple paths, which is known as multi-path propagation. Mathematically, CSI with the frequency of $f$ at time $t$ can be expressed as:
\begin{equation}
H(f, t) = \sum_{n=1}^{N} \alpha_n(t) e^{-j2\pi \frac{d_n(t)}{\lambda}},
\end{equation}
where $N$, $\alpha_n(t)$ and $d_n(t)$ are the path number, complex attenuation factor and the propagation length of the $n$-th path.
When a human moves within the range of Wi-Fi signal coverage, it causes interference to the amplitude and phase of the CSI signals, resulting in changes to the CSI signals. The changes caused by different movements vary, and we use these variations as the basis for human activity recognition.

\begin{table*}[!t]
\caption{Detailed information of  Widar3.0, XRF55, MM-Fi}
\centering

\resizebox{0.65  \textwidth}{!}{%
\begin{tabular}{c|c|c|c|c}
\hline
Dataset Name & Classes Number & Sample Number & Image Shape  & Dataset Size \\ \hline
Widar3.0 \cite{zheng2019zero} & 6              & 25014         & 22 × 20 × 20 & 5GB          \\
XRF55   \cite{wang2024xrf55}  & 55             & 33000         & 3 × 3 × 30 × 1000 & 65GB        \\
MM-Fi    \cite{yang2024mm} & 27             & 320760        & 3 × 114 × 10 & 7.5GB        \\ \hline
\end{tabular}%
}
\label{table1}
\end{table*}
\subsection{Expert Trajectory Generation}
We define the original dataset and synthetic dataset as \(D_{\text{real}}= \{(x_i, y_i)\}_{i=1}^{|D_{\text{real}}|}
\) and \(D_{\text{syn}} = \{(x_i, y_i)\}_{i=1}^{|D_{\text{syn}}|}
\)  (\(|D_{\text{syn}}| \ll |D_{\text{real}}|\)) respectively, where the data samples \(x_i, s_i \in \mathbb{R}^d\), the class labels \(y_i \in Y = \{0, 1, \ldots, C-1\}\), and \(C\) is the number of classes. Each class of \(D_{\text{syn}}\) contains \(spc\) (samples per class) data pairs, and \(|D_{\text{syn}}|  = spc \times C\).
We define expert parameters \(\theta^{D_{\text{real}}}_t\) as the teacher network parameters trained on full data at the training step \(t\), and synthetic parameters \(\theta^{D_{\text{syn}}}_t\) as the student network parameters trained on synthetic dataset. A complete expert trajectory and synthetic trajectory can be represented as \(\{\theta^{D_{\text{real}}}_t\}_{t=0}^T\) and \(\{\theta^{D_{\text{syn}}}_t\}_{t=0}^T\), where \(T\) denotes the training steps.

The essence of WiDistill lies in leveraging expert trajectories 
 \(\{\theta^{D_{\text{real}}}_t\}_{t=0}^T\) to guide the distillation of our synthetic dataset. To generate these expert trajectories, we train multiple networks on the real dataset and save their parameters at each epoch.  These parameters are termed "expert trajectories" as they represent the limit of performance achievable by a network trained on the full dataset and they can serve as expert to guide synthetic dataset to improve performance.
  We distill a dataset that induces a similar trajectory \(\{\theta^{D_{\text{syn}}}_t\}_{t=t_{0}}^T\) (starting from the same initial point \(t_{0}\)) as that induced by the real training set, resulting in a similar model.

By relying solely on real data, we can calculate the expert trajectories in advance of the distillation process. For each dataset, we conducted all experiments using this set of expert trajectories calculated beforehand. This method allows us to carry out distillation and experimentation more efficiently.

\subsection{Trajectory Matching }

In the  trajectory matching  stage, our synthetic dataset learns from the generated sequences of parameters making up our expert trajectories  \(\{\theta^{D_{\text{real}}}_t\}_{t=0}^T\). 
Our approach fosters long-term training dynamics by using our synthetic dataset to align with those observed in networks trained on real data.
At each distillation step, we first sample expert trajectory and parameters from one of our expert trajectories at a random timestep \(\theta^{D_{\text{real}}}_{t_{0}}\) and use these to initialize our student parameters \(\theta^{D_{\text{syn}}}_{t_{0}} = \theta^{D_{\text{real}}}_{t_{0}}\). Placing an upper bound \(T^+\) on \(t\) lets us ignore the less informative later parts of the expert trajectories where the parameters do not change much. With our student network initialized and the preliminary processing of Wi-Fi CSI data accomplished, we then perform \(K\) gradient descent updates on the student parameters with respect to the classification loss of the synthetic data:
\begin{equation}
\theta^{D_{\text{syn}}}_{t+n+1} = \theta^{D_{\text{syn}}}_{t+n} - \alpha \nabla \ell(D_{\text{syn}}; \theta^{D_{\text{real}}}_{t+n}),
\end{equation}
where  \(\alpha\) is the (trainable) learning rate used to update the student network.  Unlike in the field of dataset distillation in computer vision, the domain of Wi-Fi based human activity recognition does not require data augmentation techniques such as ZCA whitening that are commonly used in computer vision. Instead, we preprocess CSI signals by removing outliers and performing normalization to enhance data quality for subsequent analysis. 

\subsection{Distilled Dataset Fine-tuning}
Finally, in the fine-tuning stage of the distillation dataset, we adjust the distillation dataset according to the corresponding changes in the parameters on the distillation dataset, so that the performance of the distillation dataset approximates the real dataset.
We return to our expert trajectory and retrieve the expert parameters from \(J\) training updates after those used to initialize the student network, \(\theta^{D_{\text{real}}}_{t+J}\). Finally, we update our distilled signals based on the weight matching loss, which is the normalized squared L2 error between the updated student parameters, \(\theta^{D_{\text{syn}}}_{t+K}\), and the future expert parameters, \(\theta^{D_{\text{real}}}_{t+J}\):

\begin{equation}
L = \frac{\|\theta^{D_{\text{syn}}}_{t+K} - \theta^{D_{\text{real}}}_{t+J}\|_2^2}{\|\theta^{D_{\text{real}}}_{t} - \theta^{D_{\text{real}}}_{t+J}\|_2^2}
\end{equation}

We normalize the L2 error by the distance traversed by the expert parameters, which ensures a robust signal even during the later training epochs when the expert parameters exhibit minimal movement. This normalization also aids in calibrating the magnitude differences across neurons and layers. Although we experimented with alternative loss functions, such as cosine distance, we empirically found that the straightforward L2 loss performed better. We then minimize this objective to update the pixels of our distilled dataset and the trainable learning rate \(\alpha\) by back-propagating through all \(K\) updates to the student network. Optimizing the trainable learning rate \(\alpha\) automatically adjusts for the number of student and expert updates (hyperparameters \(J\) and \(K\)). We use SGD with momentum to optimize \(D_{\text{syn}}\) and \(\alpha\) according to this objective. Algorithm 1 illustrates our main algorithm.

\section{experiments}

\begin{table*}[!h]
\caption{Overall Performance on the Widar3.0, XRF55, MM-Fi using different numbers of distilled samples}
\centering
\resizebox{0.7\textwidth}{!}{%
\begin{tabular}{llllllll}
\toprule
  dataset &               spc &                   Ours & Kmeans & Kcenter & Herding & Rrandom & Whole Data \\
\midrule
    &  10  &                \textbf{0.3907} & 0.3898 &  0.2648 &  0.2653 & 0.3155 &            \\
 Widar3.0\cite{zheng2019zero} &  50  &               \textbf{0.6423}  & 0.4768 &  0.3595 &  0.3799 & 0.4820 &      0.9271 \\
 & 100  &                  \textbf{0.6865} & 0.5501 &  0.4361 &  0.4432 & 0.5407 &            \\
  &        10  &                 \textbf{0.2348} & 0.2197 &  0.1927 &  0.1718 &       0.1941 &            \\
 XRF55 \cite{wang2024xrf55}&        50 &                  0.5736 & \textbf{0.6395} &  0.5181 &  0.3204 &      0.5006  &       0.8905 \\
  &       100  & 0.7124  & \textbf{0.7603} &  0.6520 &  0.4103 &   0.6807 &            \\
  &       10  &  \textbf{0.1841 } &  0.1682  &  0.0507  &  0.0473  &  0.1970  &   \\
MM-Fi \cite{yang2024mm}&       50  &  0.1779  &  \textbf{0.2485}  &  0.1783  &  0.0634  &  0.1815  &  0.3402 \\
  &      100  &  \textbf{0.2809}  &  0.2706  &  0.2485  &  0.1012  &  0.2782  &   \\
\bottomrule
\end{tabular}
\label{tab2} %
}
\end{table*}

\subsection{Datasets}

Wi-Fi datasets present challenges due to their varied data types, uncertain dimensions, and high resolution, making it difficult to directly apply distillation methods developed for natural images. Therefore, there is an urgent need for a comprehensive benchmark specifically for Wi-Fi datasets to evaluate the performance of dataset distillation techniques.

We have selected some of the most recent large-scale Wi-Fi datasets,  Widar3.0\cite{zheng2019zero}, XRF55\cite{wang2024xrf55}, and MM-Fi\cite{yang2024mm}, with detailed information shown in Table \ref{table1}. These datasets encompass a variety of Wi-Fi CSI signals and their associated analysis tasks, representing the forefront of current Wi-Fi datasets. We aim to uncover the applicability and performance differences of various methods on Wi-Fi data, providing valuable benchmarks and insights for future research.


\textbf{Widar3.0} is a comprehensive Wi-Fi sensing dataset tailored for gesture recognition, comprising 22 categories and 43,000 samples. The data is collected using the Intel 5300 Network Interface Card (NIC) with 3 × 3 antenna pairs across numerous distinct environments.

\textbf{XRF55} is the Wi-Fi dataset with the most action types.
It has 55 categories, including five major categories of actions: Human-Object Interaction, Human Interaction, Fitness, Body Motion, and Human-Computer Interaction. The organizers recruited 39 subjects and asked them to repeat each action 20 times in the perception area.

\textbf{MM-Fi} consists of 27 action categories which include 14 daily activities and 13 rehabilitation exercises. The daily activities are geared towards potential smart home and building applications, while the rehabilitation categories are designed to contribute to healthcare applications.





\subsection{Overall Performance}

We generated distilled datasets with samples per class (spc) of 10, 50, and 100 respectively. Widar3.0 is trained and tested using MLP, while XRF55 and MM-Fi are trained and tested using ResNet-18. All experiments were conducted using the PyTorch framework with an NVIDIA RTX 3090 GPU with 24GB memory. Our method achieved SOTA performance overall compared to conventional methods such as Kmeans\cite{krishna1999genetic}, Kcenter\cite{khuller2000fault}, and Herding\cite{raafat2009herding}. The performance of the generated distillation dataset was tested on MLP, CNN, and ResNet, and the results are shown in Table \ref{tab2}. 

Each row in the table represents the accuracy of networks trained with the same spc extracted from different methods within the same dataset. It can be observed that, in the vast majority of cases, the networks trained on our distilled dataset achieve significantly higher accuracy than those trained using traditional methods. Specifically, when the number of spc is small, our method consistently achieves the highest accuracy. As SPC increases, only in a few cases does our method perform slightly worse than traditional methods. This is mainly because the data type of Wi-Fi CSI signals differs significantly from that of images. For instance, the shape of the data in XRF55 is (270, 1000), while images in typical computer vision tasks have shapes like (32, 32) or (64, 64)\cite{cazenavette2022dataset}. The dataset distillation algorithm alters sample values through gradient descent, which causes the changes to be more pronounced when spc is small. However, as the number of generated samples increases, the impact of distillation diminishes, thereby reducing the advantage of our methods.

\subsection{Performance On Cross-Architecture Generalization}
\begin{table}[!h]

\centering
\caption{Cross-Network Performance on Widar3.0,  XRF55, MM-Fi}
\label{table3}  
\begin{tabular}{lccc}
\toprule
       & \textbf{Ours\_CNN} & \textbf{Ours\_ResNet} & \textbf{Ours\_MLP} \\
\midrule
\textbf{CNN}     & 0.5393 & 0.5356 & 0.5434 \\
\textbf{ResNet}  & 0.5373 & 0.5693 & 0.5475 \\
\textbf{MLP}     & 0.3694 & 0.3684 & 0.4300 \\
\bottomrule
\end{tabular}
\label{tab:model_comparison}
\end{table}

\begin{table}[ht]
\centering
\begin{tabular}{lccc}
\toprule
       & \textbf{Ours\_CNN} & \textbf{Ours\_ResNet} & \textbf{Ours\_MLP} \\
\midrule
\textbf{CNN}     & 0.5785 & 0.5449 & 0.6089 \\
\textbf{ResNet}  & 0.5528 & 0.5175 & 0.5650 \\
\textbf{MLP}     & 0.5898 & 0.5393 & 0.6423 \\
\bottomrule
\end{tabular}
\label{tab:cross_network_performance}
\end{table}
\begin{table}[!h]
\centering
\begin{tabular}{lccc}
\toprule
       & \textbf{Ours\_CNN} & \textbf{Ours\_ResNet} & \textbf{Ours\_MLP} \\
\midrule
\textbf{CNN}     & 0.2020 & 0.2302 & 0.1808 \\
\textbf{ResNet}  & 0.2554 & 0.1747 & 0.2660 \\
\textbf{MLP}     & 0.1544 & 0.0879 & 0.1620 \\
\bottomrule
\end{tabular}

\end{table}
To validate the cross-domain performance of the distilled dataset generated by our method, we conducted cross-network experiments on three different datasets.  As shown in Tables \ref{table3}, for example, in the Widar3.0 dataset, we tested the distilled dataset with CNN and MLP, which was trained using ResNet. The results showed no significant drop in accuracy, and similar patterns were observed in other cases as well. 

\section{Conclusion}
Wi-Fi based human activity recognition has the advantages of protecting people's privacy and penetrating obstacles, making it highly promising in home environments. However, the increasing scale of datasets has become a major challenge. We propose the WiDistill method. Unlike traditional machine learning methods, WiDistill innovatively adopts a dataset distillation approach based on  trajectory matching, which improves the performance of the generated datasets by inducing the distilled dataset's parameter states to  align with those of the real dataset. This is the first time dataset distillation has been applied in the field of Wi-Fi based human activity recognition. We evaluated WiDistill on several large-scale Wi-Fi datasets, and it demonstrated excellent performance across the board. WiDistill offers a promising solution to the issue of excessive dataset sizes, thereby enhancing the practicality and generalizability of Wi-Fi sensing technology.

\bibliographystyle{IEEEtran}
\bibliography{reference.bib}

\end{document}